# Gas Gain Measurements from a Negative Ion TPC X-ray Polarimeter

Z. R. Prieskorn, J. E. Hill, P. E. Kaaret, J. K. Black, and K. Jahoda

*Abstract*–Gas-based time projection chambers (TPCs) have been shown to be highly sensitive X-ray polarimeters having excellent quantum efficiency while at the same time achieving large modulation factors. To observe polarization of the prompt X-ray emission of a Gamma-ray burst (GRB), a large area detector is needed. Diffusion of the electron cloud in a standard TPC could be prohibitive to measuring good modulation when the drift distance is large. Therefore, we propose using a negative ion TPC (NITPC) with Nitromethane ($CH_3NO_2$) as the electron capture agent. The diffusion of negative ions is reduced over that of electrons due to the thermal coupling of the negative ions to the surrounding gas. This allows for larger area detectors as the drift distance can be increased without degrading polarimeter modulation. Negative ions also travel ~200 times slower than electrons, allowing the readout electronics to operate slower, resulting in a reduction of instrument power.

To optimize the NITPC design, we have measured gas gain with SciEnergy gas electron multipliers (GEMs) in single and double GEM configurations. Each setup was tested with different gas combinations, concentrations and pressures: P10 700 Torr, Ne+$CO_2$ 700 Torr at varying concentrations of $CO_2$ and Ne+$CO_2$+$CH_3NO_2$ 700 Torr. We report gain as a function of total voltage, measured from top to bottom of the GEM stack, and as a function of drift field strength for the gas concentrations listed above. Examples of photoelectron tracks at 5.9 keV are also presented.

*Index Terms*-X-ray polarimetry, negative ion time projection chamber, x-ray detectors.

## I. INTRODUCTION

X-RAY polarization measurements were first reported for an astronomical source in 1972. Novick et al., [1] measured the Crab Nebula with a Bragg polarimeter on a sounding rocket flight. Weisskopf et al. [2] confirmed the results with a Bragg polarimeter flown on OSO-8 in 1976 and measured the Crab Nebula to have a polarization of 19.2% ± 1.0% at a position angle of 156 degrees. In the 30 years since, there have been few new measurements. A number of dedicated instruments were planned and some even built, e.g. SXRP [3], [4], but none have been flown. Dean et al., [5] recently reported polarization of the Crab Nebula at higher energies, 0.1 – 1 MeV with the INTEGRAL/SPI instrument. The observations were followed up with analysis of INTEGRAL/IBIS data at 200 keV [6], confirming the findings. Laurent et al., recently used the INTEGRAL/SPI instrument to make another polarization measurement. They observed the black hole X-ray binary Cygnus X-1 in the energy range 250keV-2 MeV. Using spectral modeling they resolved weakly polarized emission from 250-400 keV dominated by Compton scattering and strongly polarized emission from 400keV-2 MeV probably related to a jet [7]. The INTEGRAL results indicate the potential for X-ray polarimetry to produce new science; however, INTEGRAL is not a dedicated polarimeter and is unable to achieve polarization measurements at the 1% level due to instrument systematic effects.

Traditional methods for X-ray polarimetry have been either Bragg reflection or Thomson (Compton) scattering devices. Both of these techniques have limitations, Bragg reflectors have a very narrow energy band and scattering devices become much less efficient at energies <10 keV due to photoelectric absorption. Making use of the photoelectric effect to measure X-ray polarization with a gas proportional counter was first proposed by Austin and Ramsey in 1992 [8]. This instrument used the light generated by the electron avalanche to image the photoelectric track with a CCD camera, allowing determination of the initial direction of the photoelectron which is preferentially related to the electric field vector of the X-ray. The probability distribution of photoelectron emission angles is described by $sin^2\theta\ cos^2\varphi$, where $\theta$ is the angle of emission with respect to the direction of X-ray propagation and $\varphi$ is the X-ray electric field dependent azimuthal angle.

Costa et al. [8] constructed a gas proportional counter with pixel readout to image the electrons liberated by collisions of the photoelectron with gas atoms along its path. In this instrument, X-rays enter the detector parallel to the electric field in the drift region, which transfers the primary electrons generated along the photoelectron track from the detection region to the readout pixels. To increase quantum efficiency the detector depth along the path of the incident photon must be increased. This also increases the drift distance of the charge, increasing electron diffusion and reducing modulation. A detector with incident radiation parallel to the drift electric field can not improve quantum efficiency and modulation without making a tradeoff between the two.

To improve upon this design, Black et al. [10], suggest using a time projection chamber (TPC) for measuring X-ray

This work was supported in part by NASA grants NNX08AF46G and NNX07AF21G.

Z. R. Prieskorn performed this work while at the University of Iowa, Iowa City, IA 52240 USA (current contact information: telephone: 319-400-1809, e-mail: prieskorn@psu.edu).

J. E. Hill is with the NASA Goddard Space Flight Center, Greenbelt MD 20771 USA (telephone: 301-286-0572, e-mail: joanne.e.hill@nasa.gov).

P. E. Kaaret is with the University of Iowa, Iowa City, IA 52240 USA (telephone: 319-335-1985, e-mail: philip-kaaret@uiowa.edu).

J. K. Black is with Rock Creek Scientific, 1400 East-West Hwy, Suite 807, Silver Spring MD 20910 USA (e-mail: kevin.black@nasa.gov).

K. Jahoda is with the NASA Goddard Space Flight Center, Greenbelt MD 20771 USA (telephone: 301-286-3527 e-mail: keith.jahoda@nasa.gov)

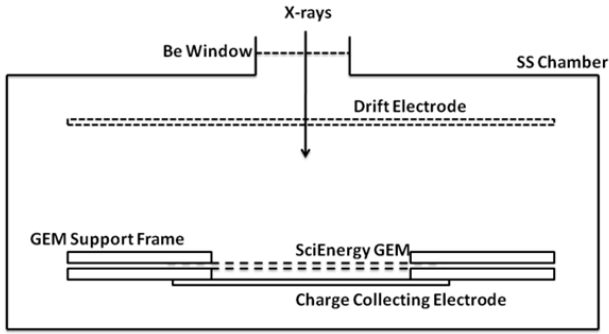

Fig. 1. GEM Testing Chamber. Used for characterizing gas electron multipliers (GEMs) and for establishing operating parameters for the negative ion time projection chamber (NITPC). The chamber is stainless steel with a ceramic insert separating the two ends. The detector is mounted on the bottom flange and a Be window on the top. The detector has a drift electrode 9 mm above the top electrode of the GEM. A collecting plate is located 1.5 mm below the bottom electrode of the GEM.

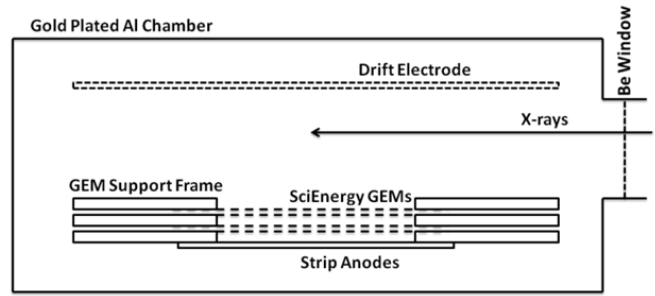

Fig. 2. NITPC Prototype detector. A prototype negative ion time projection chamber (NITPC) X-ray polarimeter. The chamber is gold plated aluminum with a stainless steel flange. A drift electrode is located 9 mm above the top electrode of the GEM. Anode strips for charge collection are located 1 mm below the GEMs bottom electrode. Used for testing single and double GEM configurations. In the double GEM configuration the distance between GEMs is 1.3 mm

polarization. The TPC polarimeter is a gas proportional counter that uses a gas electron multiplier (GEM) to produce the required electron gain [11], and strip electrodes to collect the charge. An X-ray enters the detector perpendicular to the drift electric field and interacts with a gas molecule, producing a photoelectron that is ejected in a direction preferential to the direction of the electric field of the photon. The photoelectron scatters through the gas, losing energy along its path and liberating further electrons. These electrons then drift through the uniform electric field between the drift electrode and GEM to the holes of the GEM where a Townsend avalanche occurs, multiplying the input electrons up to $10^5$ times. The charge cloud is then collected on strip anodes located beneath the GEM. The strips are instrumented with charge-sensitive pre-amplifiers and continuously sampling analog-to-digital converters (ADCs). As the charge is collected on the strips, it is read out and binned according to arrival time and strip number. By using the strip number for one dimension and the arrival time multiplied by the drift velocity for the orthogonal dimension the photoelectron track can be imaged. The emission angle is determined for each event and then binned to produce a modulation curve. Black et al. [10] give a more detailed description of the TPC X-ray polarimeter. The angle representing the maximum of the modulation curve indicates the polarization angle and the polarization magnitude is given by $\mu_m/\mu_{100}$ where $\mu_m$ is the measured source modulation and $\mu_{100}$ is the detector modulation measured from a 100% polarized source.

The TPC is able to attain good modulation and good quantum efficiency at the same time because the drift direction is perpendicular to the direction of the incoming X-ray. The quantum efficiency is improved by increasing the depth of the detector along the direction of the incident photon while modulation is optimized with a decreased drift distance.

A TPC with a small drift distance behind X-ray mirrors is excellent for measuring the X-ray polarization of existing sources, but the discovery of new transient sources, like gamma-ray bursts (GRBs), will require an instrument with a large field of view. However, as distance between drift plane and GEM increases, diffusion of the electron cloud becomes a limiting factor and will degrade the modulation. Decreasing the diffusion of the charge cloud then becomes critical to producing a large area detector. A longer drift distance allows diffusion of the charge cloud to become prohibitive to photoelectron track reconstruction. A negative ion charge cloud will be thermally coupled to the surrounding gas, the diffusion of the negative ion charge cloud is an order of magnitude less than an electron cloud [12]. Nitromethane ($CH_3NO_2$) has an electron affinity and, if introduced to the detector gas, will capture the free electrons liberated along the photoelectron path. Miyamoto et al. had success using CS2 as an electronegative gas additive but the sulfur edge prevents that from being a viable option for the 2-10 keV energy range [13]. These negative ions then drift in the uniform electric field to the higher field region of the GEM holes. The electric field is strong enough inside the GEM holes to cause collisional detachment of the ions [14]. The liberated electrons are then multiplied and detected as described for the electron TPC. Using a NITPC, large area instruments are more practical due to the improved diffusion characteristics and the reduced power requirement from slower electronics.

## II. EXPERIMENTAL SETUP FOR GEM AND NITPC TESTS

Three experimental setups for measuring gas gain were used. The GEM Testing Chamber was used for the initial single GEM gain measurements of different GEMs in different gas mixtures. In this configuration the detector is not an X-ray polarimeter, only a gas proportional counter. A second detector setup functioned as a single GEM NITPC polarimeter prototype. The third setup was a double GEM NITPC polarimeter. GEMS from several manufacturers were considered for these experiments but the availability, stability and reliability of the SciEnergy GEMS led to their being used.

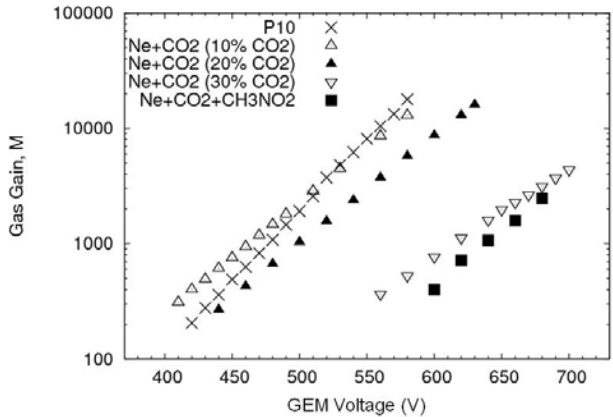

Fig. 3. Gas gain for the 3x3 cm$^2$ SciEnergy GEM. GEM voltage is the voltage between the top and bottom electrode of the GEM. GEM gain is calculated as described in §IIA. All gas mixtures were at 700 Torr total pressure. Ne+CO$_2$+CH$_3$NO$_2$ had partial pressures: Ne 600 Torr, CO$_2$ 80 Torr, CH$_3$NO$_2$ 20 Torr.

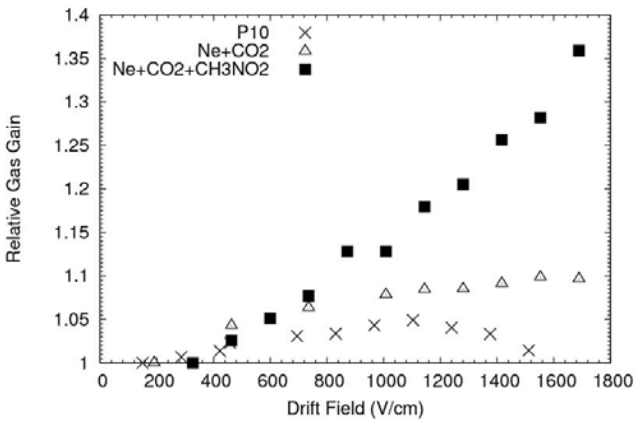

Fig. 4. A comparison of measured gain for different gases as a function of drift electric field strength. The gain is measured relative to the gain at the lowest drift field for each gas. All gas mixtures were at 700 Torr total pressure. Ne+CO$_2$+CH$_3$NO$_2$ had partial pressures: Ne 600 Torr, CO$_2$ 80 Torr, CH$_3$NO$_2$ 20 Torr.

SciEnergy GEMS are made by sandwiching a 100 μm thick liquid crystal polymer (LCP) between two layers of 5 μm thick copper. Various sizes are available with different hole patterns. The holes are cut by laser through the copper/LCP/copper layers. Holes are on a hexagonal spacing and all the GEMs in this experiment had 70 μm diameter holes on a 140 μm diameter spacing, see Tamagawa et al. for more information on SciEnergy GEMs, [15, 16].

*A. Experimental Setup of the GEM Testing Chamber*

Fig. 1 shows a schematic of the detector setup. The GEM Testing Chamber is a cylindrical vacuum chamber with a ceramic insert separating two stainless steel halves of the chamber. Each end of the chamber has a stainless steel flange mounted to it. One flange has a 125 μm thick Be window for allowing low energy X-rays to enter the chamber, 70 % transmission at 3 keV and 98 % at 10 keV [17]. The other flange has mounting points for the detector assembly, pump out and fill ports and four high voltage (HV) electrical feedthroughs.

The detector consists of a circular Al foil, thin enough for X-rays to pass through, held in an Al frame as a drift electrode, a GEM located 9 mm below the drift electrode and held in place with a plastic (delrin or PEEK) frame, and a stainless steel plate 1.5 mm below the GEM bottom layer (anode) to collect charge in place of strip anodes. X-rays are absorbed in the gas between the drift and the GEM and the charge track generated by the photoelectron drifts to the GEM where it is multiplied. We measured charge from the cathode (top layer) of the GEM to characterize GEM performance with gain curves. The cathode charge is read out through an Ortec 142-PC charge sensitive pre-amplifier, Ortec 671 shaping amplifier, using a shaping time of 10 μs due to the slow drift velocity of negative ions, and an Amptek MCA-8000 multi-channel analyzer (MCA). Amptek's PMCA software was used to read out the MCA and produce a spectrum. To prevent condensation of the CH$_3$NO$_2$, the chamber was kept between 30 - 35 °C using heating strips during operation.

*B. Experimental Setup of the Prototype NITPC X-ray Polarimeter*

A schematic of the NITPC X-ray polarimeter is shown in Fig. 2. The vacuum chamber is a rectangular gold plated aluminum body on a custom stainless steel flange. The flange has a welded 50 pin sub-D connector providing a feedthrough for the strip signals as well as four HV electrical feedthroughs. An X-ray window of 125 μm thick Be is epoxied into one of the short sides of the rectangular chamber.

The physical setup of the detector is similar to the GEM Testing Chamber but the NITPC has anode strips instead of a collecting plate and the X-rays enter the detector parallel to the plane of the GEM. A solid stainless steel plate was used as a drift electrode. A SciEnergy 2x5 cm$^2$ GEM with 70 μm holes on a 140 μm hexagonal spacing was used. The anode strips 60 μm wide copper deposited on an FR4 substrate on a 120 μm spacing, with every 24[th] strip tied to a common lead. This allows the detector to be operated with a limited electronic chains, 24 instead of hundreds.

The cathode of the GEM is instrumented exactly as in the GEM Testing Chamber, described in §IIA, except the signal from the shaping amplifier is sent to a timing discriminator. The output from the discriminator is used to trigger the readout electronics for the strips. The charge is then binned by strip and time. The drift velocity of the detector is used with the timing information and the strip position to make a 2D plot of the photoelectron track, examples of 5.9 keV tracks in Ne+CO$_2$+CH$_3$NO$_2$ are shown in Fig. 6.

*C. Double GEM Setup*

After testing the NITPC X-ray polarimeter prototype with a single GEM and finding that the gas gain was insufficient to successfully image tracks, gain measurements were made with a double GEM setup. A second SciEnergy 2x5 cm$^2$ GEM was attached to the frame used to hold the first GEM in place. A separate frame was then placed over the second GEM. The GEM holes were not aligned to each other. The spacing between GEMs was 1.3 mm.

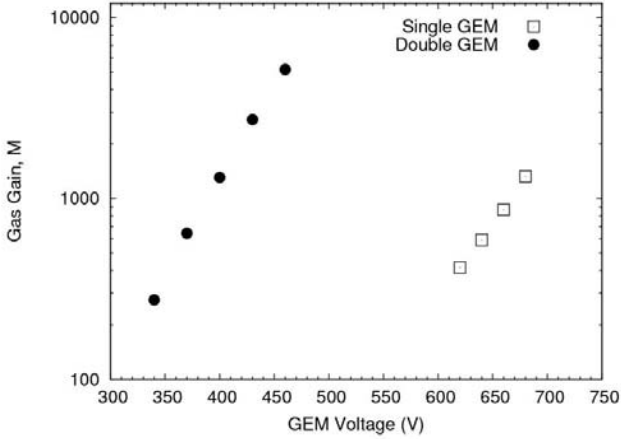

Fig. 5. A comparison of the gains measured with single and double GEM setups. The GEM voltage is the voltage across each GEM in the double GEM setup. The gas mixture for this comparison was 700 Torr of Ne+CO$_2$+CH$_3$NO$_2$ with partial pressures: Ne 600 Torr, CO$_2$ 80 Torr, CH$_3$NO$_2$ 20 Torr.

## III. RESULTS

### A. Single GEM Gain Measurements

Fig. 3 shows the approximate gas gain as a function of GEM voltage, the voltage difference between GEM cathode and anode. The GEM Testing Chamber was tested by irradiating it with X-rays from a $^{55}$Fe source. Multiple gases were tested including P10 (90% Ar, 10% methane), Ne+CO$_2$ with varying concentrations of CO$_2$ and Ne+CO$_2$+CH$_3$NO$_2$ with partial pressures of 600:80:20 Torr. All gas mixtures had a total pressure of approximately 700 Torr.

The gas gain, $M$, was obtained by using the following relation:

$$M = \frac{CVw}{eE} \frac{m}{m_o} \quad (1)$$

where $C$ is the capacitance of the charge-sensitive amplifier feedback capacitor, $V$ is the test pulse amplitude, $w$ is the average secondary ionization energy (values for different gas mixtures were taken from, [18]) $e$ is the electron charge, $E$ is the X-ray energy, $m$ is the centroid obtained from the MCA and $m_o$ is the centroid of the test pulse in the MCA, [19].

Previous TPC tests indicated that to observe tracks with sufficient signal to noise ratio to make a good estimate of the initial direction, the strip anodes need to measure a gain of at least 2000. Gas gains of 2000 can only be reached in a single GEM setup with a GEM voltage of approximately 680 V. Even with a GEM charge throughput of 100%, the GEMs would need to operate dangerously close to their breakdown voltage, found to be ~700 V, to achieve the necessary gain. Table 1 summarizes the gain measurements results.

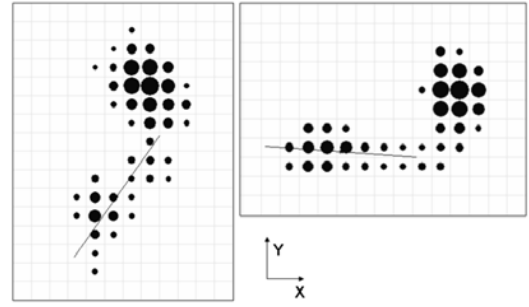

Fig. 6. Two photoelectron tracks produced with the NASA GSFC negative ion TPC (NITPC) Prototype. The Y-axis corresponds to the strip number. The X-axis is the time bin size multiplied by the drift velocity of the negative ions through the drift region.

The plot in Fig. 4 shows how the relative gain changes as a function of drift-field strength for different gases. The gain for P10 and Ne+CO$_2$ increases by a maximum of 5 and 10 %in response to drift field increases of 1.1 and 1.5 kV/cm, respectively. The curve for Ne+CO$_2$+CH$_3$NO$_2$ shows a more significant increase of 36% for a drift field increase of 1.7 kV/cm and no indication of the expected rollover in the relative gain as the electric field increases. This test was limited to a maximum drift field of 2 kV/cm due to constraints in the experimental set-up. The number of primary electrons reaching the GEM holes, via negative ion transport, is changing with drift field strength, not the gas gain itself. This could be due to electron dissociation of the negative ion or simply a contaminant removing electrons from the negative ion. Either could produce the observed effect because a higher drift field means a higher drift velocity which means less time for the removal of primary electrons resulting in a larger measured gain.

Electrostatic simulations of the drift electric field indicated that they were approximately 10 % uniform in the NITPC prototype, suggesting that electric field transport could explain the linear increase in gain with drift field. Results from a similar detector setup with 1 % uniform electric fields exhibit the same linear gain increase however, making this explanation unlikely.

Measurements of the diffusion coefficients would help resolve this issue. The different molecular species originated in the electron dissociation process would all have different coefficients of diffusion. This should result in larger than predicted diffusion of purely CH$_3$NO$_2$ negative ions.

### B. Single GEM and Double GEM Measurements with the NITPC X-ray Polarimeter Prototype

Several different detector configurations have been tested in the NITPC prototype. Nitromethane has been found to have possible material compatibility issues with bare Al, and to be absorbed by delrin and rubber. A detector consisting only of iridite Al, PEEK, stainless steel and FR4 was assembled and gain measurements were made for single and double GEM configurations. Photoelectron track images from the double GEM setup are also reported.

The lack of sufficient gas gain observed in the experiment described in §IIIA inspired the use of a double GEM setup. Fig. 5 shows gain as a function of GEM voltage for single and double GEM setups in the NITPC. The individual GEM voltages are 420 volts with the double GEM setup and 1000 V/cm between the GEMs. With this detector we were able to achieve the required gas gain of approximately 2000.

Initial photoelectron tracks measured with the double GEM NITPC are shown in Fig. 6. The source was 5.9 keV X-ray's produced from an $Fe^{55}$ radioactive source. The initial direction of the track can be clearly measured and the end and beginning of the track clearly distinguished. These observations pave the way for measuring track length distributions and polarization.

## IV. CONCLUSIONS

### A. Gain Data

The results obtained with the GEM Testing Chamber indicated that the required gains to operate a NITPC X-ray polarimeter could not be achieved with a single GEM. A NITPC X-ray polarimeter was then tested in single and double GEM configurations. With a double GEM setup, enough gas gain was achieved to produce photoelectron tracks of 5.9 keV X-rays. It is also important to note that the double GEM setup did not result in any apparent distortion of the photoelectron track. The baseline gas mixture based on results with the Single GEM Testing Chamber was for $Ne+CO_2+CH_3NO_2$ with partial pressures 600:80:20 Torr. This pressure is based on quantum efficiency and further testing with the NITPC has indicated that a lower total pressure produces better track images at this energy with reduced quantum efficiency as a tradeoff, further work is required in this area.

Charge collection when using nitromethane as a charge carrier is strongly affected by drift field strength. Further testing to find the optimum operating drift field for the NITPC and to understand the processes involved with drifting negative ions will be needed. Measuring the diffusion of the gas mixtures in this study will be an important next step in the development of this detector.

TABLE 1
GAIN RESULTS IN DIFFERENT GAS MIXTURES

| Gas Mixture | Turn on Voltage | GEM Voltage at 2000 Gain | Change in Measured Gain with Drift Field |
|---|---|---|---|
| **P10** | | | |
| 630 T Ar 70 T $CH_4$ | 420 | 500 | 5% at 1.1 kV/cm |
| **$Ne+CO_2$** | | | |
| 630 T Ne, 70 T $CO_2$ | 410 | 500 | 10% at 1.5 kV/cm |
| 560 T Ne, 140 T $CO_2$ | 440 | 530 | - |
| 490 T Ne, 210 T $CO_2$ | 560 | 650 | - |
| **$Ne+CO_2+CH_3NO_2$**[a] | | | |
| Single GEM | 600 | 670 | 36% at 1.7 kV/cm |
| Double GEM | 315[b] | 420[b] | 39% at 1.7 kV/cm |

[a] 600 T Ne, 80 T $CO_2$, 20 T $CH_3NO_2$ for the Single and Double GEM
[b] This is the voltage on each GEM in the double GEM setup

### B. Future Use for a NITPC X-ray Polarimeter

NASA recently selected the Gravity and Extreme Magnetism SMEX (GEMS) as one of two new small explorer missions. GEMS will measure X-ray polarization in the 2-10 keV energy range from accreting black holes, magnetars and supernovas.

GEMS will be great for known sources but will leave a gap in observing polarization from transient sources like GRBs, SGRs and transient black holes. A polarimeter based on the NITPC could be used to make these measurements. A 2 year mission with a modest instrument (6 kg, 6 W) could potentially measure polarization from 12 GRBs [20]. Polarization measurements of the prompt X-ray emission could identify the emission mechanism for this radiation.

ACKNOWLEDGMENT

We would like to thank C.J. Martoff and Michael Dion for their help with this work. The Brookhaven National Laboratory National Synchrotron Light Source was used for a portion of the data in this experiment, we would like to thank Syed Khalid in particular for his help at BNL. We would also like to thank Israel Moya, Christian Urba, Richard Koenecke and Tracy L Pluchak-Rosnak for their technical expertise.